\documentclass[12pt]{article}
\usepackage{graphicx}

\newsavebox{\sboxpubnumber}
\newsavebox{\sboxpubdate}
\newcommand{\pubdate}[1]{\begin{lrbox}{\sboxpubdate}{#1}\end{lrbox}}
\newcommand{\pubnumber}[1]{\begin{lrbox}{\sboxpubnumber}{\begin{tabular}{l} #1 \\
				 \usebox{\sboxpubdate}
				 \end{tabular}}
                           \end{lrbox}
                           \pubblock}
\newcommand{\Title}[1]{\begin{center} {\Large #1 } \end{center}}
\newcommand{\Author}[1]{\begin{center}{ \sc #1} \end{center}}
\newcommand{\Address}[1]{\begin{center}{ \it #1} \end{center}}

\newcommand{\pubblock}{\rightline{
			\usebox{\sboxpubnumber}}}
\newenvironment{Abstract}{\begin{quotation}  }{\end{quotation}}
\newenvironment{Presented}{\begin{quotation} \begin{center}
             PRESENTED AT\end{center}\bigskip
      \begin{center}\begin{large}}{\end{large}\end{center}
      \end{quotation}}

\begin{document}

\begin{titlepage}
\pubdate{\today}                    
\pubnumber{XXX-XXXXX \\ YYY-YYYYYY} 

\vfill
\Title{Archeops: CMB Anisotropies Measurement from Large to Small Angular Scale}
\vfill
\Author{Alexandre Amblard, on behalf of the Archeops Collaboration
}
\Address{PCC, Coll\`ege de France  \\
         Paris, France}
\vfill
\begin{Abstract}

Archeops, a balloon-borne experiment, will provide a measurement of
CMB anistropies from large to small angular scale thanks to its large
sky coverage (30\%), its high angular resolution (10 arcminutes), and
its high signal-to-noise ratio due to high sensitivity 100 mK cooled
bolometers. We will therefore be able to put strong constraints on the
value of the cosmological parameters. Archeops flew already twice,
once in Sicily for a technical flight in 1999, and once from Sweden
for its first scientific flight in January 2001. I describe here
Archeops'main characteristics, the preliminary results from the
scientific flight, the expected precision of this flight for power
spectrum measurements, and perspectives for the next flights this
winter.

\end{Abstract}
\vfill
\begin{Presented}
    COSMO-01 \\
    Rovaniemi, Finland, \\
    August 29 -- September 4, 2001
\end{Presented}
\vfill
\end{titlepage}
\def\thefootnote{\fnsymbol{footnote}}
\setcounter{footnote}{0}

\section{Introduction}
The Cosmic Microwave Background (CMB) was discovered by Penzias and
Wilson \cite{PenziasWilson:1965} in 1965, and interpreted by Dicke et
al. \cite{Dickeetal:1965}. This radiation comes from the first moments
of our Universe, and was first predicted by Gamow, Alpher and Hermann
\cite{AlpherHermGamow:1948} in 1948, in the context of the Big Bang
theory. The Big Bang theory predicted that our Universe was in
expansion, which was first observed by Hubble \cite{hubble:1929} in
1929, via the redshift of galaxies. As a consequence of the expansion,
the Universe also cools down, meaning that at earlier times it was
hotter. Going back in the past, our Universe was so hot that matter
and photons were tightly coupled with each other. They formed a plasma
in thermal equilibrium. As the Universe cooled down, the photon energy
became too small to ionise the matter, below 0.1 eV (below 13.6 eV due
to the high photon to electron ratio), the mean free path of the
photons became larger than the horizon. The photons were free to cross
the Universe towards our detectors. The blackbody dsitribution and
spatial properties of these photons remain unchanged due to their
negligeable cross-section with matter, the blackbody temperature is
just cooled to 2.7 K due to the Universe expansion. Through this
radiation, we obtain a picture of our Universe 300 000 years after the
Big Bang.

\section{Motivations}
\begin{figure}[h!]
    \centering \includegraphics[height=3.5in]{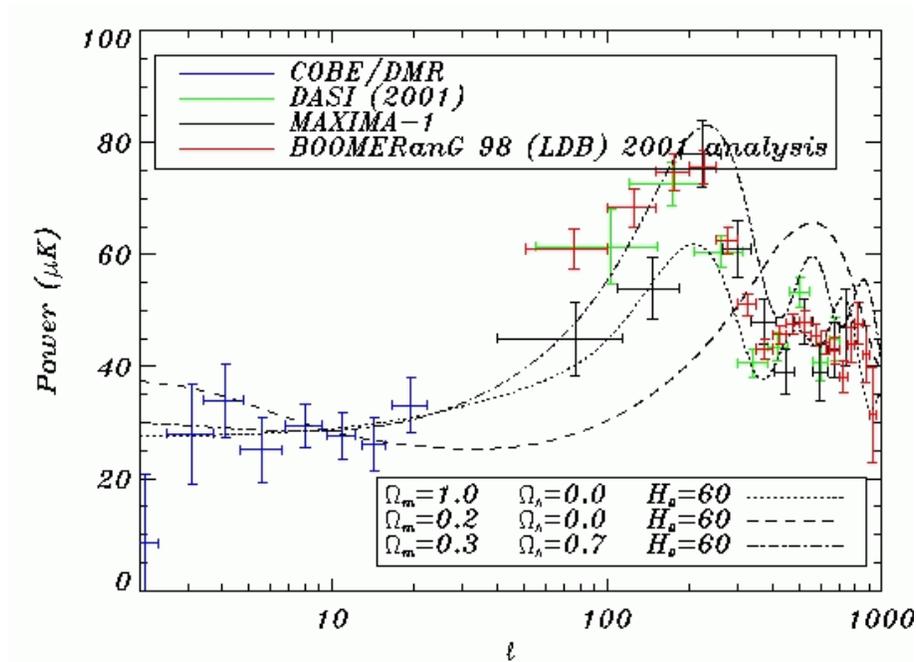}
    \caption{Most accurate available results, taken by the COBE
    satellite, and the experiments Boomerang, Maxima, DASI. We note
    the lack of measurements between COBE and other experiments.
    Archeops'main goal is to link these data sets with one single
    measurement. Three different inflationnary models are also
    plotted, showing the influence of the cosmological paremeters on
    the power spectrum.}
\label{fig:recentresults}
\end{figure}

\noindent
After the extremely accurate CMB spectrum measurement by the COBE
satellite in 1989, proving that the radiation was a perfect blackbody
\cite{matheral:1994,fixsenal:1996}, and the first detection of
deviations \cite{smoothal:1992} from the perfect homogeneity at the
level of 10$^{-5}$ K, most experiments have concentrated on improving
the measurement of the anisotropies. In most theories and with
consistent available data, the statistical distribution of the CMB
anistropies is Gaussian. The anisotropies are therefore represented by
their spherical harmonic power spectrum (non-Gaussianity is
nevertheless looked for). The most accurate results are from COBE on
large angular scale ($>$ 7 degrees) and on small angular scales ($<$ 2
degrees) from Boomerang \cite{netterfieldal:2001}, Maxima
\cite{hanany:2000}, and DASI \cite{halverson:2001} (see Figure
\ref{fig:recentresults}). The intermediate angular scales, between
COBE and the other experiments, lack measurements, due to the small
sky coverage of these experiments. The main Archeops goal is to link
COBE and Boomerang-Maxima-DASI angular scales, with high-sensitivity
detectors covering a large fraction of sky (30\% of the sky). Archeops
is also a testbed for Planck-HFI, using the same type of cryogenic
system and detectors; and it measures dust and galactic sources in a
polarized frequency band.

\section{Instrument and Scan-strategy}
%

\begin{figure}[h!]
    \centering \includegraphics[height=3.5in]{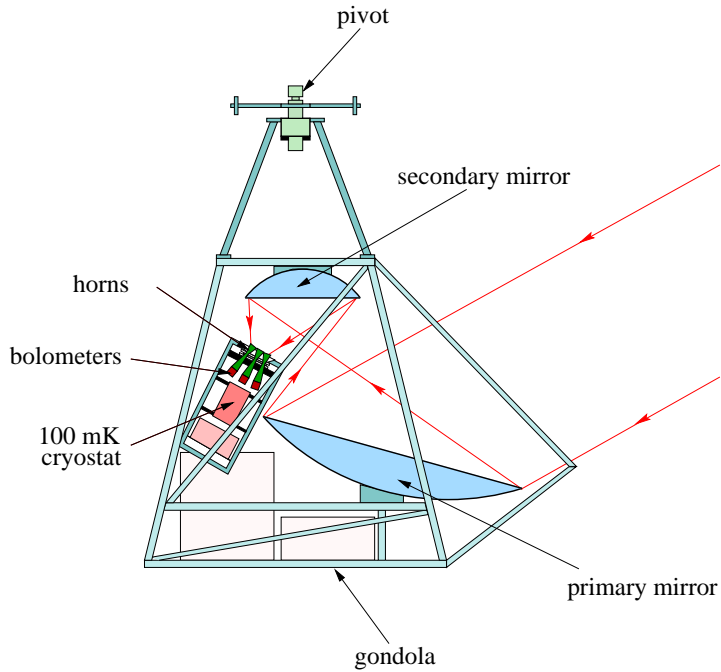}
    \caption{Schematic drawing of the gondola: the off-axis gregorian
    telescope, the cryostat, and the bolometers with their feeding
    horns. The pivot on the top of the gondola spins the payload.}
\label{fig:gondolaschema}
\end{figure}

A complete description of Archeops can be found in \cite{benoit:2001} or on 
http://www.archeops.org, but the main characteristics of Archeops are:
\begin{itemize}
\item {high angular resolution (10 arcminutes), due to a large 
primary mirror of 1.5 meters and to our horns, which guide the
radiation to our bolometers;}
\item {30\% sky coverage due to a scan-strategy consisting 
of making large circles on the sky. Associated with the first
characteristic, this allows Archeops to measure the CMB power spectrum
from angular scales of about 20 degrees (l=30) down to 10 arcminutes
(l=800);}
\item {bolometers cooled to 100 mK, by the same type 
of cryogenic system which will be used by Planck HFI, and does not
rely on gravity;}
\item  {a frequency band, the 353 GHz band, polarized using OMT, 
is dedicated to the measurement of Galactic dust and point sources
polarization.}
\end{itemize}
The Archeops telescope is an off-axis gregorian telescope (see Figure
\ref{fig:gondolaschema}), pointed at 41 degrees elevation, which defines 
the size of our circles on the sky at about 200 degrees.

\begin{figure}[h!]
    \centering \includegraphics[height=3.5in]{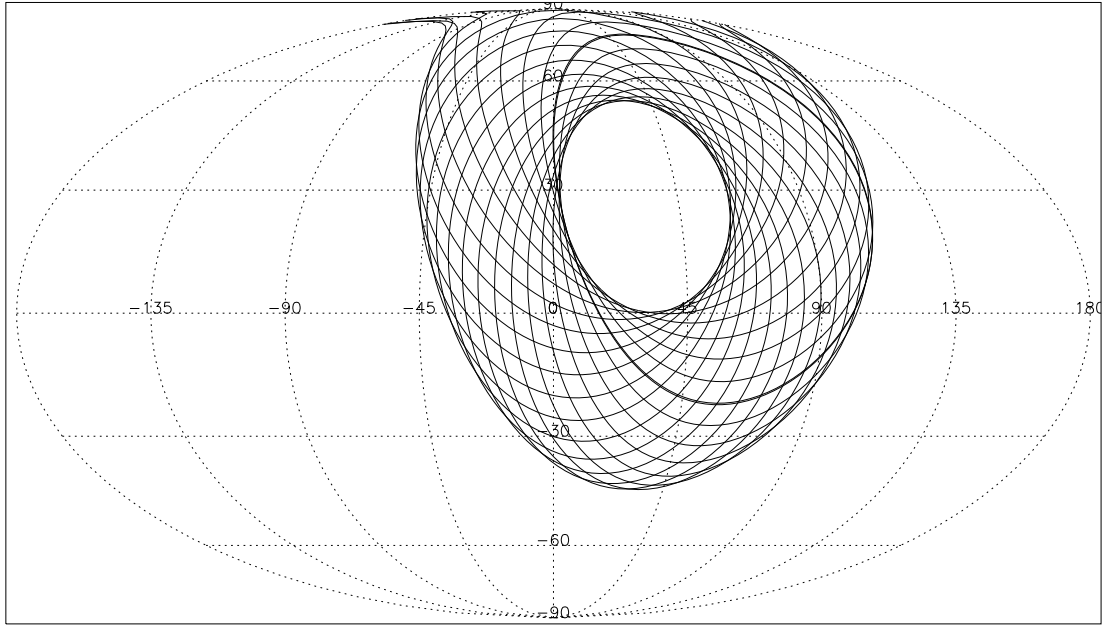}
    \caption{Typical Archeops sky coverage : each line represents a
    circle, the time separation between the circles is 1 hour.}
\label{fig:donutcoverage}
\end{figure}

\noindent
To make circle on the sky the gondola spins at constant elevation, and
as the earth rotates, the center of the circle describes also a circle
on the sky. This describes a typical ring or donut-like sky coverage
(see Figure \ref{fig:donutcoverage}). For Archeops (41 degrees
elevation and 68 degrees of latitude), the sky coverage is about 30\%
for a 24 hour flight.

\begin{figure}[h!]
    \centering 
    \includegraphics[height=2.in]{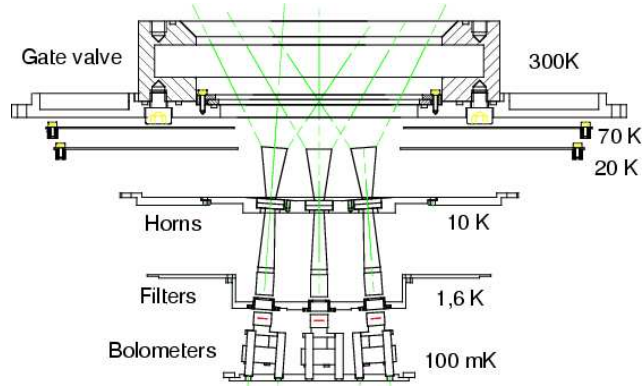}
    \caption{Optical path after the secondary mirror, from the
    membrane to the bolometers. The light passes through the horns and
    filters, to be received by the bolometers.}
\label{fig:horn}
\end{figure}

\noindent
After the secondary mirror, the radiation goes through a polypropylene
membrane (see Figure \ref{fig:horn}), and enters the horns, which
define the part of the sky seen by each bolometer; they play a crucial
role for the angular resolution of the experiment.

\begin{figure}[h!]
    \centering{ \includegraphics[height=2in]{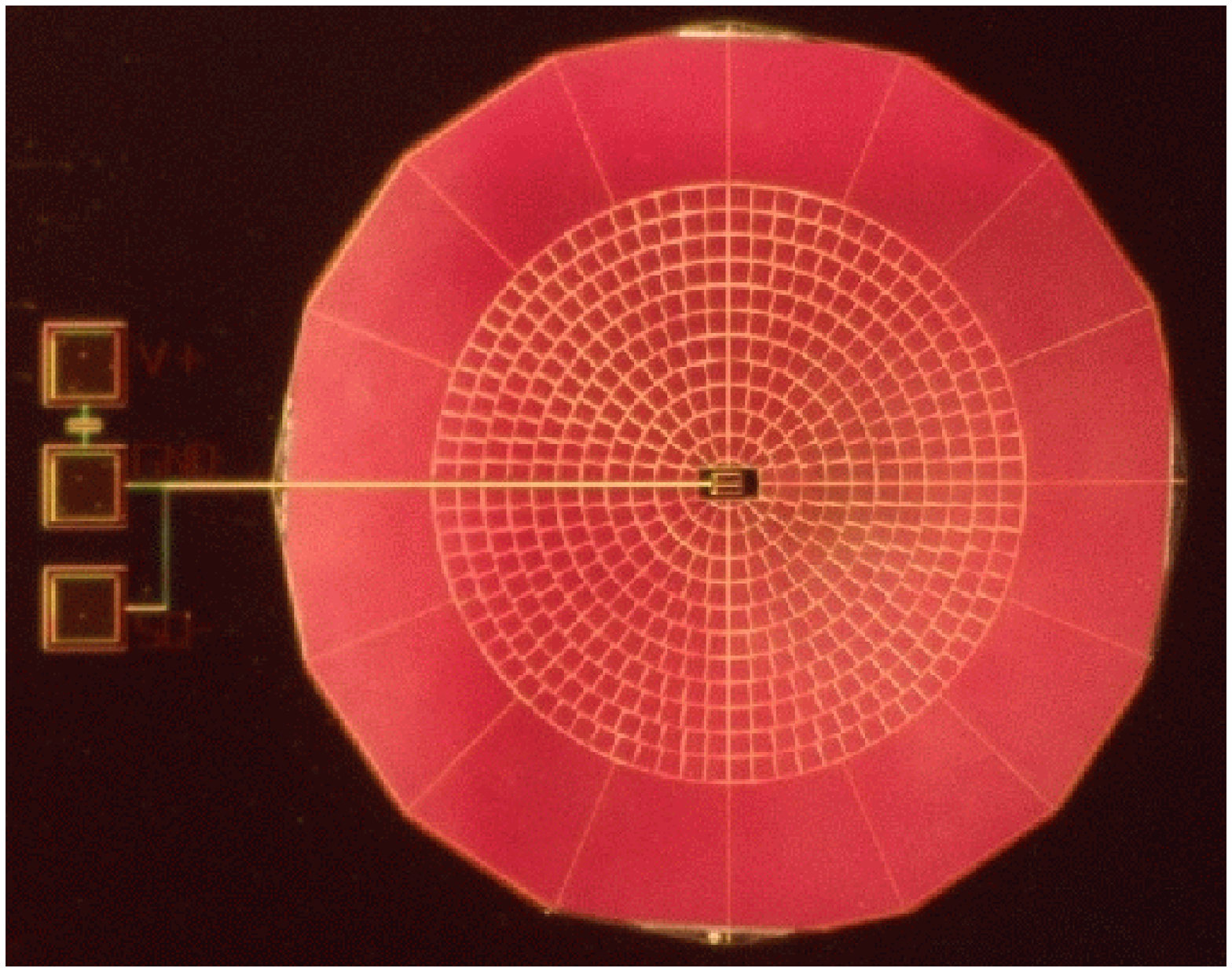}
    \includegraphics[height=2in]{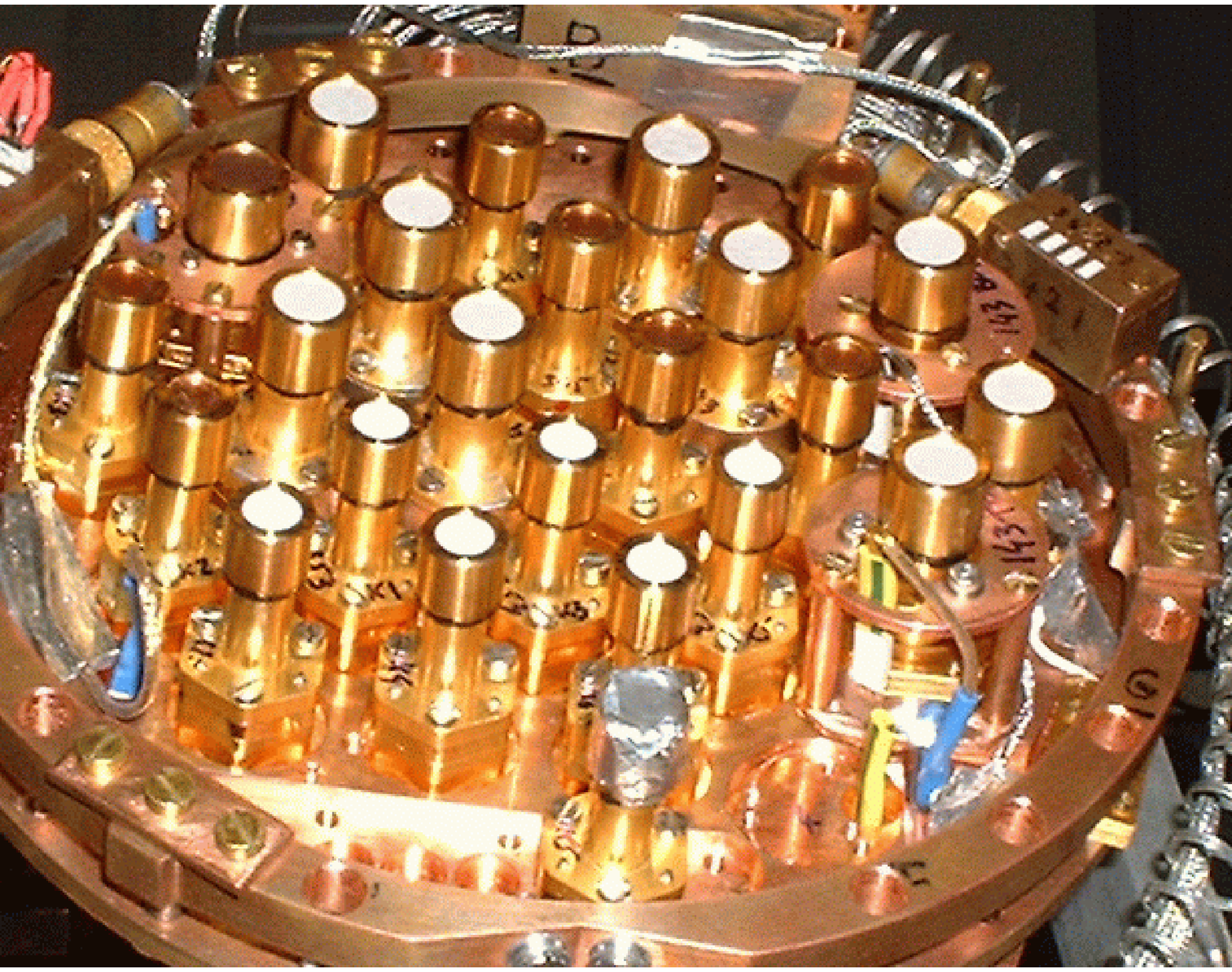}} 
   \caption{Spider web bolometer (left) with mesh size about 1 mm, the Archeops focal
    plane (right) filled with its 22 bolometers. This stage
    corresponds to the lower (100 mK) stage of Figure \ref{fig:horn}.}
\label{fig:spiderweb}
\end{figure}

\noindent
The detectors are spiderweb bolometers (Figure \ref{fig:spiderweb}),
which have a small calorific capacity so a fast response time, and are
less sensitive to cosmic ray hits. These bolometers are cooled to 100
mK by a $^3$He-$^4$He dilution, which is produced in capillary tubes
placed around the focal plane (see Figure \ref{fig:spiderweb}). The focal
plane is filled with 22 bolometers, with the following frequency band
distribution :
\begin{itemize}
\item {8 detectors at 143 GHz, where the CMB is the dominant emission
 at large galactic latitude;}
\item {6 detectors at 217 GHz, where the CMB is still dominant but dust 
contamination is larger;}
\item {6 detectors at 353 GHz, where the dust and atmospheric emission 
are dominant; these 6 channels are polarized using OMT : 3 pairs of 2
detectors share the same horn, the signal being separated into 2
orthogonal polarized components;}
\item {2 detectors at 545 GHz, where dust and atmospheric emission 
are dominant.}
\end{itemize}

\begin{figure}[h!]
    \centering{ \includegraphics[height=3in]{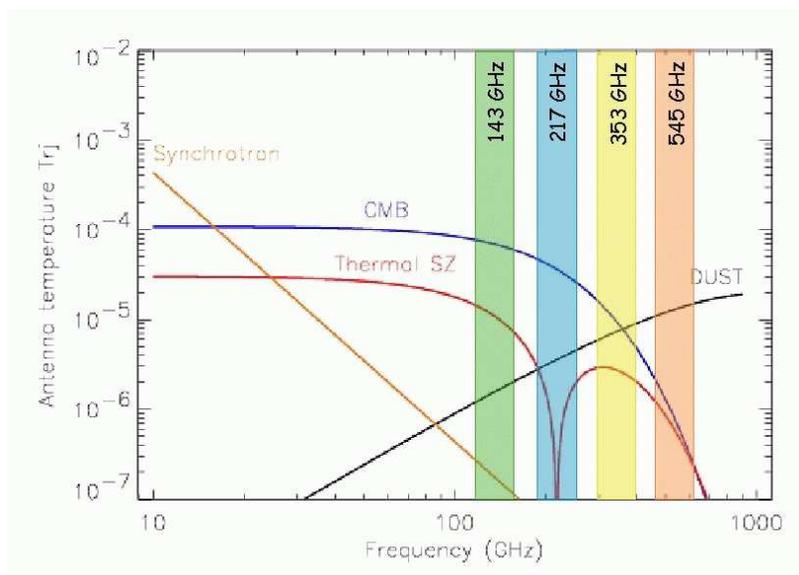}}
    \caption{Frequency bands of Archeops, with the spectrum of
    different astrophysical processes, estimated around a galactic
    latitude of 70 degrees, and a galactic longitude of 340 degrees
    (adapted from \cite{delabrouille:2001}). The different bands
    allow us to separate different emission sources, as they have
    different spectral shape.}
\label{fig:bandemission}
\end{figure}

\noindent 
This frequency distribution is used to distinguish the different
contributions of astrophysical origin or from parasitic signals, such
as atmospheric emission; this is illustrated in Figure
\ref{fig:bandemission}.

\section{Flights}

The Archeops experiment has already flown twice, once for a
technical flight launched in Trapani (Sicily, Italy) crossing the
mediterranean sea towards Spain in July 1999, producing 4 hours of night
data, but with only 4 detectors. The second flight started from Esrange,
a SSC (Swedish Space Corporation) base, near Kiruna (Lapland in the
north of Sweden). The gondola was launched by the CNES (Centre
National d'Etudes Spatiales\footnote{The French National Space
Agency}) on January, 29$^{th}$ 2001 at 2pm (local time) and landed in Russia
at Syktyvkar around midnight.

\noindent
The flight was shortened due to strong stratospheric winds (about
400km/h), which pushed the gondola too rapidly towards the east. We
however obtained 7.5 hours of scientific data (during night and at a
float altitude of 31.5 km), covering 22.7\% of the sky. The cryostat
remained below 100 mK during the entire flight.

\begin{figure}[htb]
    \centering \includegraphics[height=4.in]{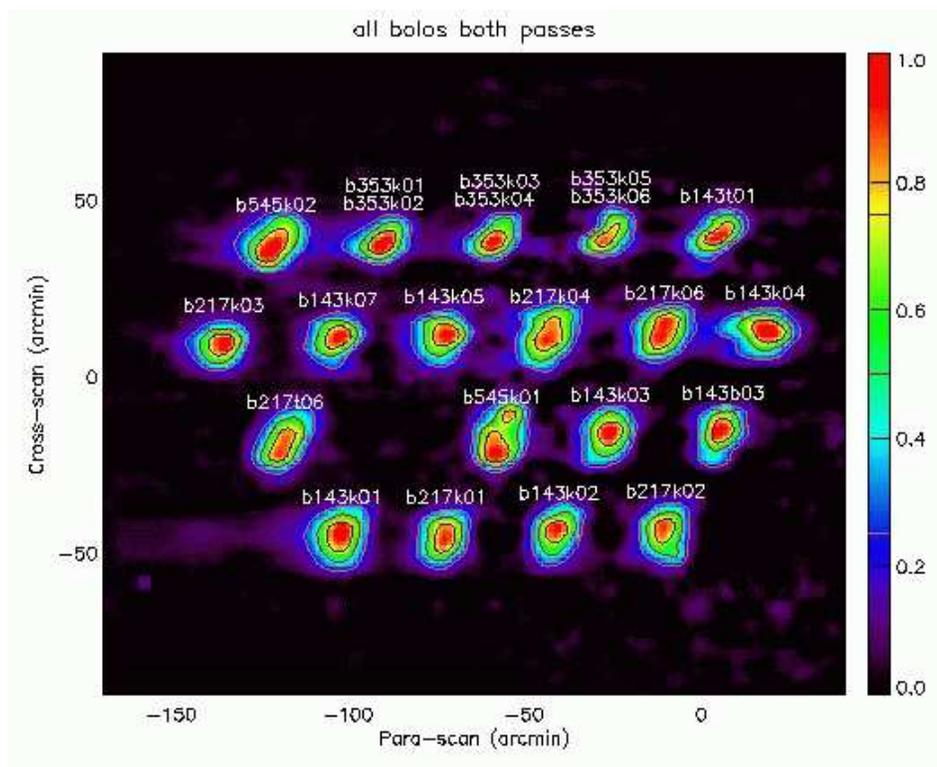}
    \caption{Archeops beams (angular sensitivity distribution),
    projected on the sky, obtained using Jupiter's signal.}
\label{fig:beams}
\end{figure}

\noindent
The average size of our optical beam is around 8 arcminutes, the
average effective beam size is around 12 arcminutes because of the
bolometers time response. Some beams are also a bit asymetric, but
this only impacts the very smallest angular scales, not attainable by
Archeops for this flight (see Figure \ref{fig:estps10detkir}). The
quality of the data was excellent, and will allow us to compute an
accurate CMB anisotropies spectrum, especially on the low $\ell$ edge
of the first acoustic peak.

\section{Expected performance and future prospects}
\begin{figure}[h]
    \centering \includegraphics[height=3.5in]{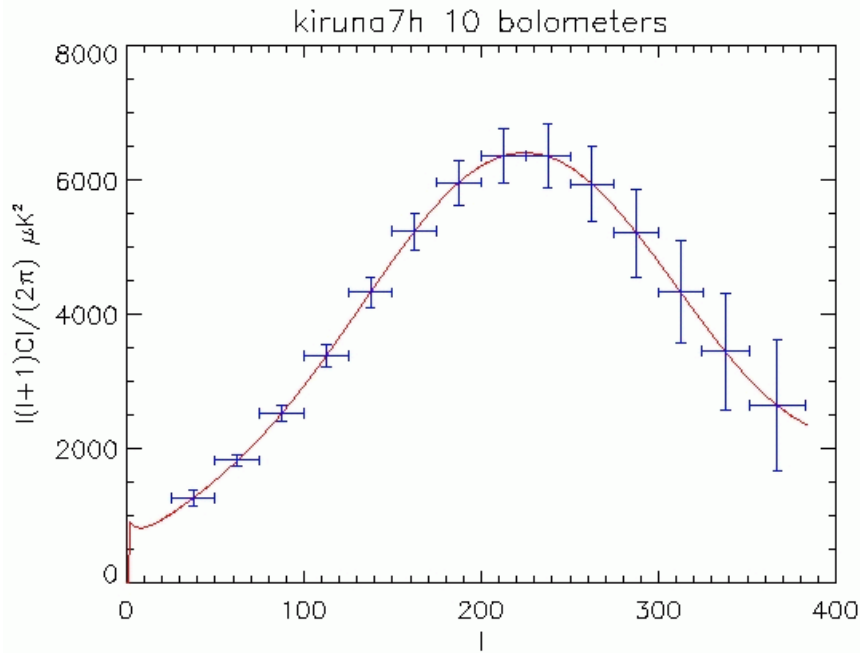}
    \caption{Power spectrum estimated from simulations, using 10
    detectors with our best noise level, and averaging their power
    spectra using their correlation matrices. The estimation is centered
    on the input model as it is the average of one thousand realizations,
    showing that the method is unbiased. Error bars are very small at
    large angular scale before the first acoustic peak at l=200 due to
    our large sky coverage, but increase quite fast afterwards on small
    angular scale due to the short duration of our flight.}
\label{fig:estps10detkir}
\end{figure}

The data analysis is not yet finished, with various tests to check the
power spectrum estimation still to be performed. Using the complete
data processing pipeline and the known noise properties of our data,
we have performed simulations for 10 detectors, based on the noise
power spectrum (in the time domain) of our best channel. The power
spectrum estimated from these simulations is shown in Figure
\ref{fig:estps10detkir}. It is the average of power spectra given by
each detector, weighted by their correlation matrices. In addition we
performed an average over several simulations to check that the method
was not biased.

\begin{figure}[htb]
    \centering \includegraphics[height=3.5in]{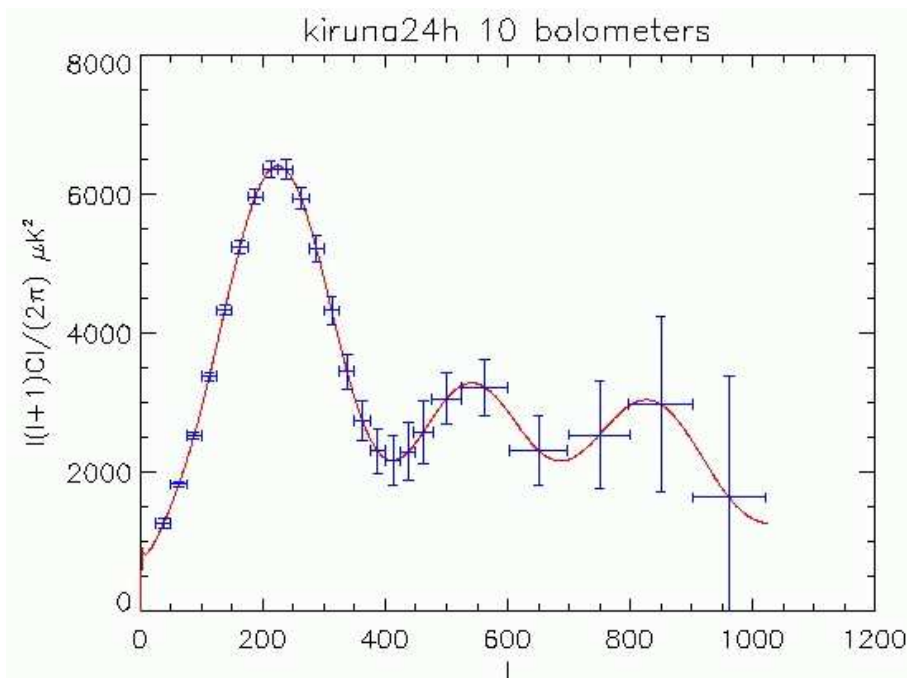}
    \caption{Power spectrum estimated from simulation : same method
    as for Figure \ref{fig:estps10detkir}, but for a 24 hours
    flight. The determination of the first peak is nearly perfect, and
    the measurement is good until the second peak.}
\label{fig:estps24hours}
\end{figure}

\noindent
The error bars are very small (smaller than available measurement) on
large angular scale, because of our large sky coverage, and then
increase after the first acoustic peak because of the small
signal-to-noise ratio due to the small integration time. This is the
motivation for a longer duration flight of at least 24 hours, as the
signal-to-noise ratio increases nearly quadratically after the first
10 hours (because the sky coverage does not increase further). Two
flights are therefore scheduled for next Winter (December 2001 -
January 2002), from Esrange, with the same configuration. The expected
performance for a 24 hour flight with 10 bolometers is shown in Figure
\ref{fig:estps24hours}.

\noindent
The power spectrum estimated this way (Figure \ref{fig:estps24hours})
is very accurate up to the end of the first peak and gives good
constraints on the second one, allowing precise estimation of
cosmological parameters.

\section{Conclusions}

Archeops is a balloon-borne experiment, designed to measure the CMB
anistropies from large (about 10 degrees) to small angular scales
(about 10 arcminutes). Archeops is also a testbed for Planck HFI, and
will provide useful information on galactic dust and point source
polarisation. The balloon has already flown twice successfully: once
during a test flight and another time for its first scientific flight
from Kiruna. The first scientific flight yielded 7.5 hours of
scientific data, with 14 detectors dedicated to CMB
measurements. Results from this flight will come soon, and we showed
that it will give good constrains on large angular scale, but less
constraining on small angular scale due to the short duration of the
flight. Archeops will fly again this Winter, in order to increase
the accuracy on large scales and measure the small scales.

\end{document}